\documentclass{aa} 

\usepackage{graphicx}
\usepackage{xcolor}
\newcommand{\red}{\textcolor{black}}

\usepackage{txfonts}
\begin{document} 
   \title{
 Properties of the cosmological filament between two clusters: \\
\textbf{A possible}
detection of a large-scale accretion shock by \textit{Suzaku}
   }

   \author{
   H. Akamatsu\inst{1}\and
   Y. Fujita\inst{2}\and
   T. Akahori\inst{3}\and
   Y. Ishisaki\inst{4}\and 
   K. Hayashida\inst{2}\and
   A. Hoshino\inst{5}\and \\
   F. Mernier\inst{1,6}\and
   K. Yoshikawa\inst{7}\and 
   K. Sato\inst{8}\and
   J. S. Kaastra\inst{1,6}
          }

   \institute{SRON Netherlands Institute for Space Research, Sorbonnelaan 2, 3584 CA Utrecht, The Netherlands
              \email{h.akamatsu@sron.nl}
         \and
        Department of Earth and Space Science, Graduate School of Science, 
	Osaka University, Toyonaka, Osaka, Japan
         \and
          Graduate School
of Science and Engineering, Kagoshima University, 1-21-35, Korimoto,
Kagoshima, Kagoshima 890-0065, Japan
          \and
          Department of Physics, Tokyo Metropolitan University, 1-1 Minami-Osawa, Hachioji, Tokyo, Japan
          \and 
          Research Center for Measurement in Advanced Science, Faculty of Science, Rikkyo University 3-34-1,  
          Nishi-Ikebukuro,Toshima-ku, Tokyo Japan
	\and
         Leiden Observatory, Leiden University, PO Box 9513, 2300 RA Leiden, The Netherlands          
          \and 
          Center for Computational Sciences, University of Tsukuba, 1-1-1 Tennodai, Tsukuba, Ibaraki, Japan
          \and
          Department of Physics, Tokyo University of Science,
1-3 Kagurazaka, Shinjyuku-ku, Tokyo, 162-8601, Japan
             }
   \date{}

 
\abstract{
We report on the results of a {\it Suzaku} observation of 
the
 plasma in the filament located  between 
the two massive clusters of galaxies Abell\,399 and Abell\,401. Abell\,399 ({\it z}=0.0724) and Abell\,401 ({\it z}=0.0737) are expected to be in the initial phase of a cluster merger.  
In
 the region between the two clusters, we find a clear enhancement in the temperature of the filament plasma from 4 keV (expected value from a typical cluster temperature profile) to {\it kT}$\sim$6.5 keV. 
 Our analysis also shows that filament plasma is present out to a radial distance of 15\arcmin~(1.3 Mpc) from a line  connecting the two clusters. The temperature profile is characterized by an almost flat radial shape with {\it kT}$\sim$6--7 keV within 10\arcmin~or~$\sim$0.8 Mpc. Across {\it r}=8\arcmin~from the axis, the temperature of the filament plasma shows a drop from 6.3 keV to 5.1 keV, indicating the presence of a shock front.  The Mach number based on the temperature drop is estimated to be ${\cal M}\sim$1.3. We also successfully determined the abundance profile up to 15\arcmin~(1.3 Mpc), showing an almost constant value ($Z$=0.3 solar) at the cluster outskirt.
We estimated the Compton {\it y}-parameter to be $\sim$14.5$\pm1.3\times10^{-6}$, which is in agreement with {\it Planck}'s results (14-17$\times10^{-6}$ on the filament). The line of sight depth of the filament is {\it l}$\sim$1.1 Mpc, indicating that the geometry of filament is likely a pancake shape rather  than cylindrical. The total mass of the filamentary structure is $\sim$7.7$\times10^{13}~\rm M_{\odot}$.
We discuss a possible interpretation of the drop of X-ray emission at the rim of the filament, which was pushed out by the merging activity and formed by the accretion flow induced by the gravitational force of the filament.
  }
  \authorrunning{H. Akamatsu et al.}
  \titlerunning{Witnessing an active structure formation cite
between two clusters of galaxies by {\it Suzaku}}

\maketitle

\section{Introduction}\label{sec:intro}
According to the scenario of hierarchical structure formation, two important predictions are made:
(i) large-scale accretion shocks located between the virial and turnaround radii of clusters of galaxies and 
(ii) about one-third of the baryons in the Universe exist as a warm-hot intergalactic medium (WHIM) in the large-scale filamentary structure of galaxies \citep{fukugita98, cen99, dave01, yoshikawa03, yoshikawa04, springel05}. In spite of their importance for  the cluster evolution and  the generation of  high energy photons and cosmic rays (CR)~\citep[e.g.,][and see references therein]{kang96, Miniati01a, Miniati01b, miniati03,ryu03, skillman08}, large-scale accretion shocks have not been detected yet because of limited capabilities of current observatories.
Studying properties of the WHIM such as density, temperature, and metallicity is essential to understand the evolution of the an inter-galactic medium (IGM) and galaxies.  The properties of the WHIM in filaments of galaxies have been investigated before.  For instance, the density of the WHIM in filaments has been deduced from observations of X-ray absorption lines \citep{takei02, fujimoto04, nicastro05, kaastra06,buote09, fang10, zappacosta10, nicastro13}. Up to now, however, X-ray emission from the WHIM has not been detected yet, due to limited sensitivities of X-ray spectrometers such as CCD cameras, except for reports of angular correlation of the diffuse X-ray background \citep{galeazzi09} and emission in a filament linking two clusters of galaxies \citep{werner08}. 
The latter case,  it is slightly denser and hotter than what expected for the WHIM, 
would be promising to observe with present-day X-ray detectors, although such a medium might be partly mixed with an intra-cluster medium (ICM).

Abell\,399 (hereafter A399) and Abell\,401 (A401) are one for the cluster associations for which a filamentary structure linking them is known.  We define the collision axis as the axis connecting the two cluster centers. Table~\ref{tab:info} summarizes basic information of these  clusters. 
Line-of-sight and projected distances between the centers of the clusters are $\sim$6 Mpc and $\sim$3 Mpc, respectively, and thus the three-dimensional distance between them is $\sim$7 Mpc.

Using HEAO1 and HEAO2 ({\it Einstein}), \cite{ulmer79} and \cite{ulmer81} first reported diffuse X-ray emission extending $\sim$2 Mpc between A399 and A401. \cite{fujita96} observed A399 and A401 with {\it ASCA}, and confirmed an excess of X-ray emission over the value expected for the case of a simple superposition of the clusters. 
They showed the presence of IGM between A399 and A401. 
These results were also confirmed by \cite{sakelliou04} with {\it XMM-Newton}. 
Later, \cite{fujita08} observed a field between both clusters with {\it Suzaku}, and found a metallicity of 0.2 solar \cite[abundance table,][]{angr89}. They claimed that the proto-cluster region was polluted with metals by superwinds from galaxies before the clusters were formed. A similar result was also reported with {\it Suzaku} observations of the Perseus cluster \citep{werner13}. More recently, {\it Planck} has detected the thermal Sunyaev-Zel'ovich (tSZ) effect \citep{planck13_filament}. Combined with {\it ROSAT} X-ray data, they derived a  temperature and an electron density of the filament of $kT$=7.1$\pm$0.9 keV, $n_e=(3.7\pm0.2)\times10^{-4}~\rm cm^{-3}$, respectively.

\citet{akahori08} performed N-body + smoothed particle hydrodynamics simulations of A399 and A401 to examine the effects of a non-equilibrium ionization state and an electron-ion two-temperature structure on the estimation of the metallicity in the linked region. Furthermore, they predicted that  shock fronts could exist offset from the collision axis. Cosmological simulations suggest that the density perpendicular to the collision axis would have an exponential profile~\citep[e.g., ][]{colberg05, dolag06}.

In this paper, we present the results of a reanalysis of the deep 150 ks {\it Suzaku} observation of the region between A399 and A401.  We assume cosmological parameters $H_0 = 70$ km s$^{-1}$ Mpc$^{-1}$, $\Omega_{\rm M}=0.27$ and $\Omega_\Lambda = 0.73$. As our fiducial reference for the proto-solar abundances  denoted by $Z_\odot$, we adopt \citet{lodders09} and Galactic absorption with $n_{\rm H}=1.0 \times 10^{21} \rm~cm^{-2}$ \citep{dickey90}. 
Unless otherwise stated, the errors shown in this paper are  68\% confidence for a single parameter.

\begin{figure}
\begin{tabular}{c}
\begin{minipage}{\hsize}
\begin{center}
  \includegraphics[width=.95\hsize]{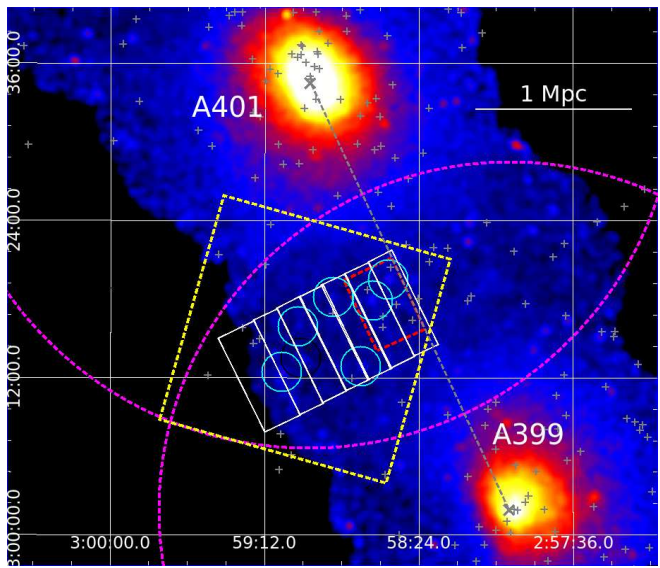}
\end{center}
\end{minipage}
\end{tabular}
\caption{
{\it XMM-Newton} image of A399 and A401 in the 0.5--2.0 keV band. 
The yellow box shows the field of view of the {\it Suzaku} XIS.
Each region is shown in the figure as a red (6\arcmin$\times$4\arcmin)and white box (2\arcmin$\times$8\arcmin or 3\arcmin$\times$8\arcmin).
Magenta circles indicate the virial radius of each cluster.
Small cyan circles show the excluded point sources.
Grey pluses represent galaxies with available redshift between $0.07<{\it z}<0.08$ in the NASA/IPAC Extragalactic Database (NED).
}
\label{fig:xmm}
\end{figure}

\begin{table}
 \centering
  \caption{Basic information of Abell 399 and Abell 401}
  \begin{tabular}{@{}llrrrrrlrlr@{}}
  \hline
Cluster & (R.A, DEC)$^a$&$z$ &$kT^b$& $r_{200}^c$  \\
	    &				&	 & keV	& Mpc   \\
 \hline
A399 & (2h57m, +13d02m) & 0.0724 & 7.23 & 2.19\\
A401 & (2h58m, +13d34m) & 0.0737 & 8.47 & 2.19\\
\hline
\multicolumn{4}{l}{$^a$: \citet{oegerle01}}\\
\multicolumn{4}{l}{$^b$: \citet{sakelliou04}}\\
\multicolumn{4}{l}{$^c$: \citet{reiprich02}}\\
\end{tabular}
\label{tab:info}
\end{table}

\section{Data reduction}
{\it Suzaku} has performed a deep 150 ks observation in the region between A399 and A401 during AO1 (from 2006-08-19, ObsID: 801020010). In this paper, we discuss the results based on {\it Suzaku} XIS data. The {\it Suzaku} XIS consists of three front-side illuminated (FI) CCD chips (XIS0, XIS2, and XIS3) and one back-side illuminated (BI) chip (XIS1)~\citep{koyama07}.  All  observations were performed with either the normal $5\times5$ or $3\times3$ clocking mode. We used FTOOLS from the HEADAS 6.12 package and CALDB 2016-06-07 for all of the {\it Suzaku} data analysis presented here. We performed event screening with the cosmic-ray cut-off rigidity COR2 $> 6$ GV to increase the signal to noise ratio.  The total exposure times after data screening is 128.3 ks.

\section[]{Spectral Analysis and Results}\label{sec:results}
\subsection[]{Spectral modeling}\label{sec:spec}
We performed spectral analysis of the {\it Suzaku} XIS data to obtain the spatial distribution and physical properties of the plasma in the filament between both clusters. The basic strategy of the fitting is similar to~\citet{akamatsu16}. We used an {\it XMM-Newton} archival observation for point source identification (green circles in Fig.~\ref{fig:xmm}).  We excluded the point sources with a flux $> 2\times 10^{-14}~\rm erg/cm^2/s$ by a 1.5\arcmin~ radius to take into account the point spread function (PSF) of the {\it Suzaku} XRT~\citep{xrt}. 

For the spectral analysis, we used the SPEX\footnote{https://www.sron.nl/astrophysics-spex}~\citep{spex} software version 3.03.00.\,  
Each spectrum was binned based on the {\tt optimal binning} method~\citep{kaastra16}. The best-fit
parameters were obtained by minimizing the C-stat.  A decrease in the low-energy efficiency of the XIS due to  contamination of the optical blocking filter was also included in ancillary response files (ARFs). 
We generated ARFs assuming uniformly-extended emission from an observed region with a 20\arcmin~radius. We carried out spectral fits to the pulse-height spectrum in each region separately.  We used the energy ranges of 0.8--8.0 keV for the both detectors.

The observed spectra consist of three different components such as (1) Non X-ray instrumental background (NXB), (2) X-ray emission from the sky objects and, (3) X-ray emission from the filament plasma.  The NXB component was estimated from the dark Earth database using the {\bf xisnxbgen} program in FTOOLS \citep{tawa08}. To adjust for the long-term variation of the XIS background due to radiation damage, we accumulated the NXB data for the period between 150 days before and 150 days after. The second background component, X-ray emission from sky objects, can be subdivided into Galactic emission and unresolved X-ray sources (Cosmic X-ray background: CXB).   These background components are modeled during the fitting procedure. We use the following model for these sky background components: $cie+abs*(cie+powerlaw)$. 
Here, the unabsorbed and absorbed {\textit cie} components represent X-ray emission from the local hot bubble (LHB) and our Galaxy (the Milky Way halo: MWH), respectively. The powerlaw component represents the contribution from unresolved point sources (CXB).
We fixed the temperature of each thermal component to 0.08 keV and 0.27 keV~\citep{yoshino09}, respectively.  The redshift and abundance of both components were fixed at zero and one solar, respectively.
We kept the normalization of thermal components free. For the CXB component, we refer to~\citet{kushino02}.
\begin{figure}
\begin{tabular}{c}
\begin{minipage}{1\hsize}
\begin{center}
\includegraphics[width=.45\hsize,angle=-90]{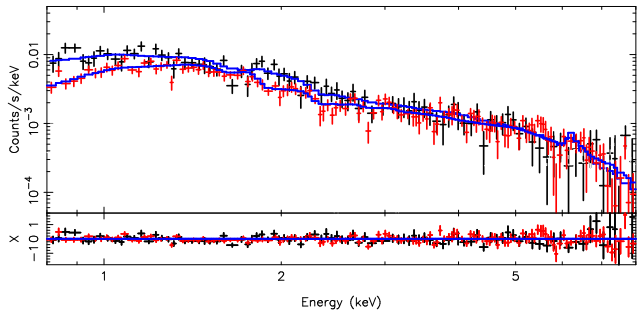}
\end{center}
\end{minipage}
\end{tabular}
\caption{\label{fig:spec}
Example of the spectral fitting. Spectra after subtraction of the NXB and the point sources. The XIS BI (Blue) and FI (Red) spectra are fitted with CXB + Galactic components (LHB and MWH) and the X-ray emission from the filamentary plasma.
}
\end{figure}

\begin{table}
\caption{Best-fit parameters for the 4\arcmin$\times$6\arcmin~box along with the collision axis \label{tab:46box}}
\begin{center}
\begin{tabular}{cccccccccccc} \hline
{\it kT} &  {\it Z} & Norm & C-stat/d.o.f. \\
(keV)	    & ({\it Z}$_\odot$) &  (10$^{69}/\rm m^3/\Box\arcmin$) \\ \hline
$  {6.52}\pm0.35 $  & $ {0.23}\pm0.08 $ & $ {23.8}\pm{0.5}$ & 377 / 299\\
\hline
\end{tabular}
\end{center}
\end{table}%

\subsection[]{Results}\label{sec:result}
\begin{figure}
\begin{tabular}{c}
\begin{minipage}{\hsize}
\begin{center}
\includegraphics[width=1.0\hsize]{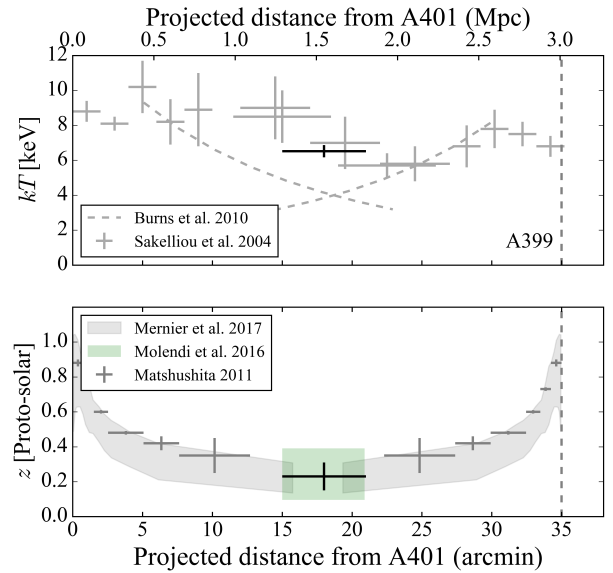}
\end{center}
\end{minipage}
\end{tabular}
\caption{\label{fig:T_comp}
Top:
Radial profiles of the temperature of the ICM and filamentary plasma. 
The {\it Suzaku} best-fit value and statistical errors are shown with the black  cross. 
The gray crosses show the previous XMM-{\it Newton} measurements by ~\citet{sakelliou04},  with converted 1$\sigma$ errors. 
The gray dashed curves show ``universal" temperature profiles \citep{burns10}. 
The gray dashed vertical line shows the approximate position of A399.
Bottom: Radial profiles of the abundance. 
All values are converted into \citet{lodders09}.
The grey shed area (scatter of the sample) and crosses represent the Fe abundance profiles reported by \citet{mernier17} and~\citet[][]{matsushita11}, respectively. Green shed area shows the conservative prediction of the abundance at cluster outskirts by~\citet{molendi16}.
}
\end{figure}

First, we investigated the properties of filamentary plasma along the collision axis. We extracted  the spectra within
 a 6\arcmin$\times$4\arcmin~(504$\times$336 kpc$^{2}$) box (red box in Fig.\ref{fig:xmm}
 and fitted them with an absorbed thin thermal plasma model ($abs\times cie$) together with the sky background components discussed above. 
The best-fit parameters and the spectral fit are shown in Table~\ref{tab:46box} and Fig.~\ref{fig:spec}, respectively.
The top panel of Fig.~\ref{fig:T_comp} displays the projected temperature, together with previous results \citep{sakelliou04}.  We also plot a ``universal" temperature profile proposed by \cite{burns10} as grey dotted curves. The authors performed N-body+hydrodynamic simulations of the cluster formation, and compared the density, temperature, and entropy profiles with those of two relaxed clusters observed with {\it Suzaku}. From the comparison, they constructed a model of the temperature profile as
\begin{equation}\label{equation:burns10}
\frac{T}{T_{\rm avg}}=A\left[1+B\left({\frac{r}{r_{200}}}\right)\right]^{\beta}.
\end{equation}
The model is in agreement with the two relaxed clusters (Abell\,1795 and PKS0745-191) observed with {\it Suzaku}, as well as other {\it Suzaku} results~\cite[see][]{akamatsu11, reiprich13}. {\it Suzaku}'s temperature estimate significantly exceeds the ``universal" profile assuming $T_{\rm avg}=7.23, 8.47$ keV~\citep{sakelliou04} and $r_{200}=21.9, 21.9$ arcmin~\citep{reiprich02} for A399 and A401, respectively.
The bottom panel of Fig.~\ref{fig:T_comp} displays the abundance profile.
We also plot a Fe abundance profiles reported by \citet{mernier17} and \cite{matsushita11} as grey shed area and grey crosses, respectively. 
Furthermore, it is worthwhile to mention that our abundance value is in good agreement with the prediction by~\citet{molendi16}, who did a conservative estimation of the Fe abundance at cluster outskirts.

\begin{table*}
\caption{
Best-fit parameters of the filamentary plasma 
\label{tab:results}
}
\begin{center}
\begin{tabular}{ccccccccc} \hline
Region$^{a}$			& Temperature	& Abundance	& $Norm$	& C-stat/d.o.f. \\ 
(arcmin)			&	(keV)	& ({\it Z}$_\odot$) 		& (10$^{69}/\rm m^3/\Box\arcmin$) \\ 
\hline
$ 1.0 \pm {1.0} $ & $ 6.43 \pm {0.87} $ & $ 0.33 \pm {0.15} $  & $ 24.3 \pm {0.6} $ & 235.60 / 207 \\ 
$ 3.0 \pm {1.0} $ & $ 6.33 \pm {0.43} $ & $ 0.23 \pm {0.10} $  & $ 21.8 \pm {0.5} $ & 302.03 / 233 \\ 
$ 5.0 \pm {1.0} $ & $ 6.37 \pm {0.47} $ & $ 0.13 \pm {0.11} $  & $ 18.4 \pm {0.5} $ & 261.96 / 223 \\ 
$ 7.0 \pm {1.0} $ & $ 6.36 \pm {0.41} $ & $ 0.33 \pm {0.10} $  & $ 16.4 \pm {0.4} $ & 254.83 / 229 \\ 
$ 9.0 \pm {1.0} $ & $ 5.91 \pm {0.42} $ & $ 0.25 \pm {0.10} $  & $ 14.5 \pm {0.4} $ & 278.48 / 231 \\ 
$ 11.0 \pm {1.0} $ & $ 5.07 \pm {0.33} $ & $ 0.31 \pm {0.11} $  & $ 13.3 \pm {0.4} $ & 253.28 / 221 \\ 
$ 13.5 \pm {1.5} $ & $ 4.86 \pm {0.27} $ & $ 0.17 \pm {0.08} $  & $ 12.9 \pm {0.3} $ & 425.51 / 276 \\ 
\hline
\multicolumn{4}{l}{$^a$: Distance from the collision axis}\\ 
\end{tabular}
\end{center}
\end{table*}%

\begin{table*}
\caption{
Best-fit parameters for inside and outside the temperature break
\label{tab:shock}
}
\begin{center}
\begin{tabular}{ccccccccc} \hline
Region$^{a}$			& Temperature	& Abundance	& $Norm$	& C-stat/d.o.f. \\ 
(arcmin)			&	(keV)	& ({\it Z}$_\odot$) 		& (10$^{69}/\rm m^3/\Box\arcmin$) \\ 
\hline
10.0--14.0 & $ 5.05 \pm {0.21} $ & $ 0.27 \pm {0.07} $  & $ 13.5 \pm {0.3} $ & 488.24 / 346 \\ 
 2.0--8.0  &   $ 6.27 \pm {0.27} $ & $ 0.28 \pm {0.07} $  & $ 18.6 \pm {0.3} $ & 369.97 / 309 \\ 
\hline
\multicolumn{4}{l}{$^a$: Distance from the collision axis}\\ 
\end{tabular}
\end{center}
\end{table*}%

\begin{figure}
\begin{tabular}{c}
\begin{minipage}{1\hsize}
\begin{center}
\includegraphics[width=1.0\hsize]{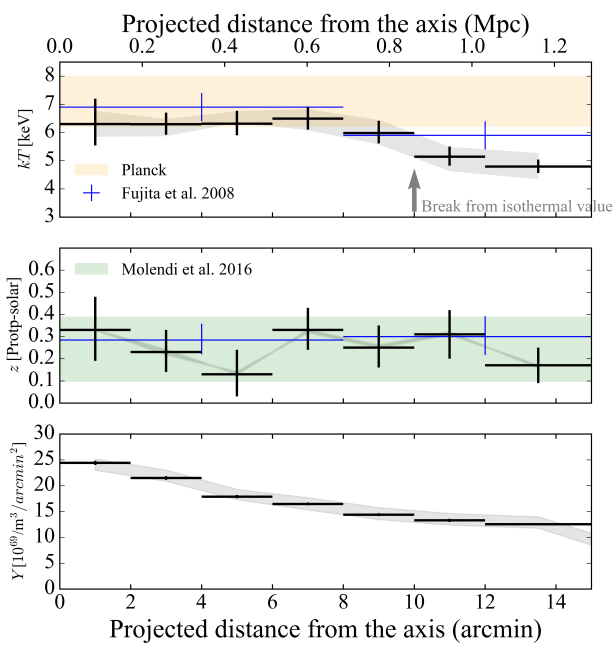}
\end{center}
\end{minipage}
\end{tabular}
\caption{\label{fig:T_pro}
Top: Profiles of the temperature of the filamentary structure as a function of the projected distance from the collision axis shown in figure~\ref{fig:xmm}. Black crosses are the Suzaku best-fit values with statistical errors. 
The blue crosses show the previous measurements by~\citet{fujita08}, with converted 1$\sigma$ errors. 
Light orange shed area represents the temperature estimated by {\it Planck}~\citep{planck13_filament}.
Middle: Radial profiles of the abundance. Colors are same as upper panel except for light green shed area, which  show the best constrained limits of the Fe abundance at cluster outskirts derived by~\citet{molendi16}.
Bottom: Same as the top panel but for the normalization of the IGM component.
The range of uncertainties due to the 30\% fluctuation in the CXB is shown by the grey shadow region. 
}
\end{figure}

\begin{figure}
\begin{tabular}{c}
\begin{minipage}{\hsize}
\begin{center}
  \includegraphics[width=.95\hsize]{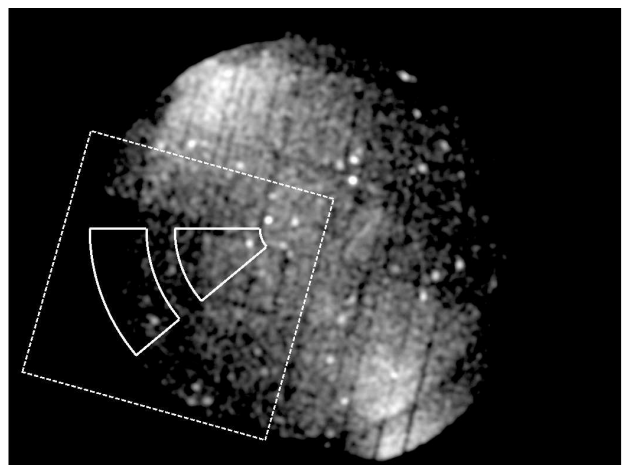}
\end{center}
\end{minipage}
\end{tabular}
\caption{
{\it XMM-Newton} image of A399 and A401 in the 0.5--2.0 keV band. 
The white dotted box shows the field of view of the {\it Suzaku} XIS.
}
\label{fig:on_filament}
\end{figure}

In order to investigate the spatial distribution of the plasma in the filament, we extracted spectra in 2 or 3$\times$8~arcmin$^2$ box regions perpendicular to the collision axis as shown in Fig.~\ref{fig:xmm}. 
We followed \red{the same spectral analysis approach}. Intending to measure emission from the space between both clusters, the redshift was fixed to {\it z}=0.073. We obtained fairly good fits for all the regions with the above model.  The best-fit values are summarized in Table~\ref{tab:results}. The radial profiles of temperature, abundance, and normalization ({\it Y}) of the {\it cie} component are shown in Fig.~\ref{fig:T_pro}. 

To investigate the robustness of the temperature measurements, we consider possible systematic errors due to the background model. 
The intensity of the filamentary plasma is much stronger than that of the background components.
In that case, the major contribution to the systematic error comes from the fluctuations of the CXB intensity. 
Here we note that our results are insensitive to the Galactic background components because their emissions are out below the energy range that was used for the spectral fitting.
In this paper, we checked the effect of the fluctuation of the CXB intensity by changing the intensity with $\pm30\%$ from the nominal value. 
In order to estimate the amplitude of the fluctuation, we followed the approach described in \citet{akamatsu11}. 
Using the observed area and the flux limit of the excluded point sources, the fluctuations were estimated to be 16--24\%. To be conservative, we adopt a fluctuation of 30\% for all regions.
The uncertainty due to the CXB fluctuations is shown as the gray shed area in Fig.~\ref{fig:T_pro}.

\begin{figure}
\begin{tabular}{c}
\begin{minipage}{\hsize}
\begin{center}
  \includegraphics[width=.95\hsize]{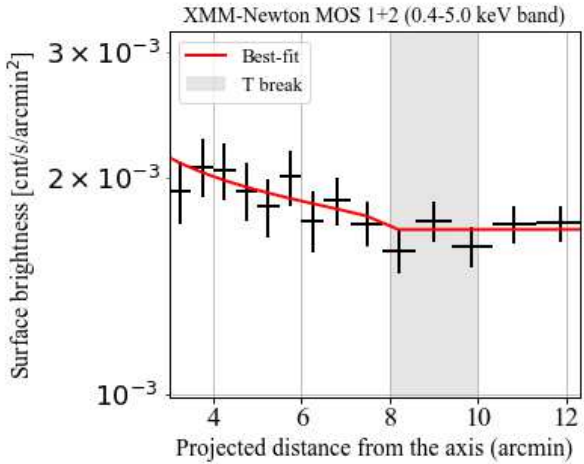}
\end{center}
\end{minipage}
\end{tabular}
\caption{
{\it XMM-Newton} surface brightness profile in the 0.4--5.0 keV band. The best-fit profile was derived by using the  \texttt{PROFFIT} package~\citep{eckert12}.
}
\label{fig:proffit}
\end{figure}

The temperature profile is almost flat at {\it kT}$\sim$6-7 keV from the collision axis up to 10\arcmin~($\sim$0.8 Mpc), which agrees well  with previous results such as {\it XMM-Newton}~\citep{sakelliou04} and {\it Suzaku} \citep{fujita08}.  In particular, {\it Suzaku}'s results show excellent agreement with {\it Planck}'s tSZ measurement~\citep[{\it kT}=7.1$\pm$0.9 keV:][]{planck13_filament}. 
Across {\it r}=8\arcmin--10\arcmin, the temperature of the filamentary plasma deviates from the isothermal value {\it kT}$\sim$6.5 keV to {\it kT}$\sim$5 keV.  
To investigate the temperature break, we extracted spectra by selecting annuli from 2.0--8.0\arcmin~and 10.0--14.0\arcmin~ from the collision axis with an opening angle of 180\degr--240\degr~(Fig.\ref{fig:on_filament}). The results are summarized in Table.~\ref{tab:shock}, which is consistent with the results of the box-shaped analysis.  Although the significance is low due to limited statistics, we also found a sign of a break in the X-ray surface brightness profile from a short XMM-Newton observation. 
We extracted the surface brightness profile from the data and made a model with a broken power-law density profile that includes a jump from a shock using the \texttt{PROFFIT} package~\citep{eckert12}. The profile and best-fit model are shown in Fig.~\ref{fig:proffit}.
Considering the rarelity of the cold front beyond 0.5 $r_{200}$ (see Fig.5 in Walker et al. 2014),  we conclude that the temperature break is a shock front.  In the next section, we discuss a possible origin of this temperature break.

The middle panel of Fig.~\ref{fig:T_pro} shows the
 radial profile of the abundance in the filament plasma.
 Thanks to the low and stable detector background of {\it Suzaku} XIS, we were able to derive abundances up to $\sim1.3$ Mpc (15 arcmin) with a reasonable accuracy.  The measured abundance values are in good agreement with the previous measurements~\citep[after considering the difference of the abundance table:][]{fujita08} and  results of the Perseus cluster~\citep{werner13}. 

\begin{figure*}
\begin{tabular}{cc}
\begin{minipage}{.5\hsize}
\begin{center}
\includegraphics[width=1.\hsize]{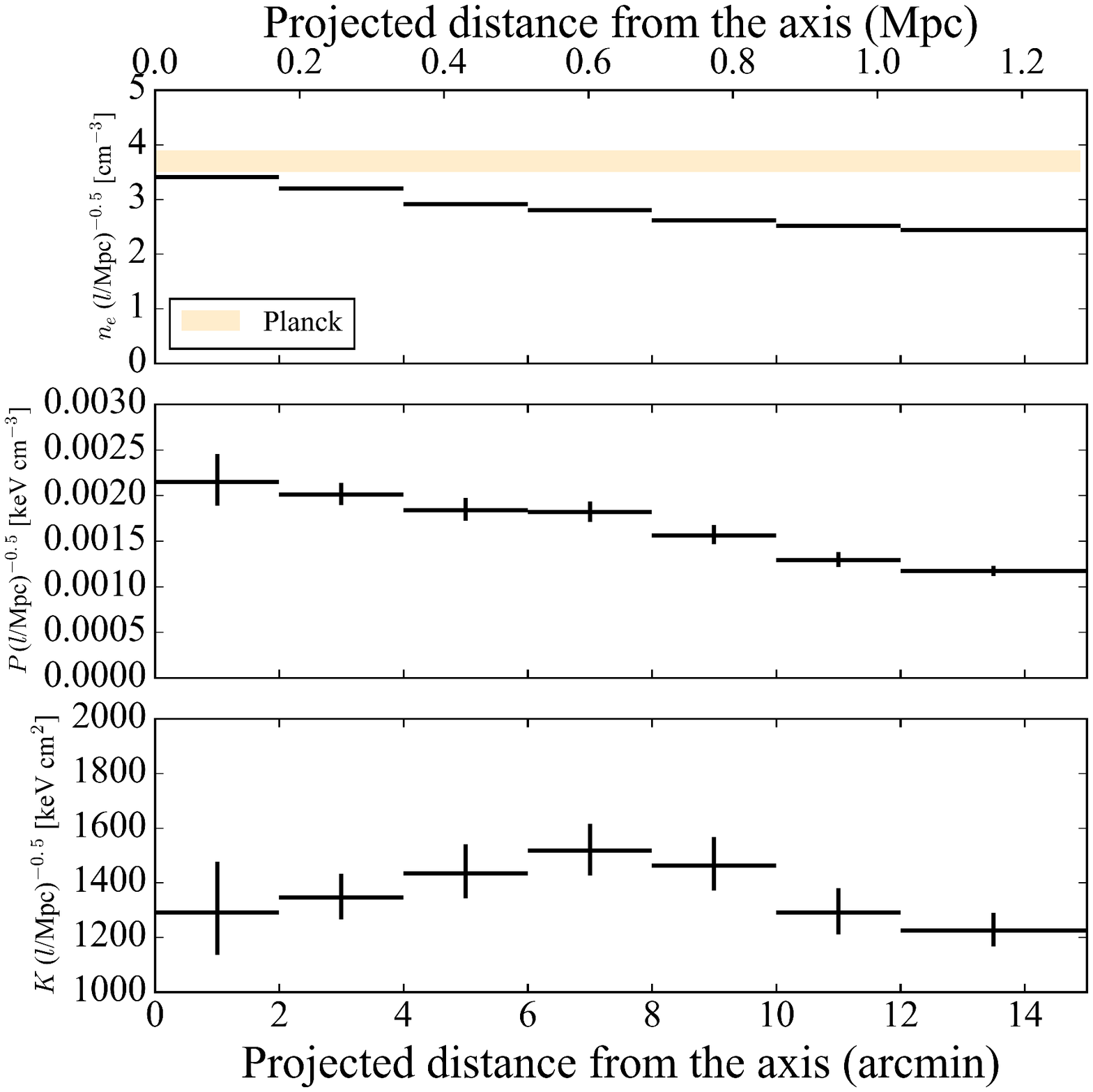}
\end{center}
\end{minipage}
\begin{minipage}{.5\hsize}
\begin{center}
\includegraphics[width=1.\hsize]{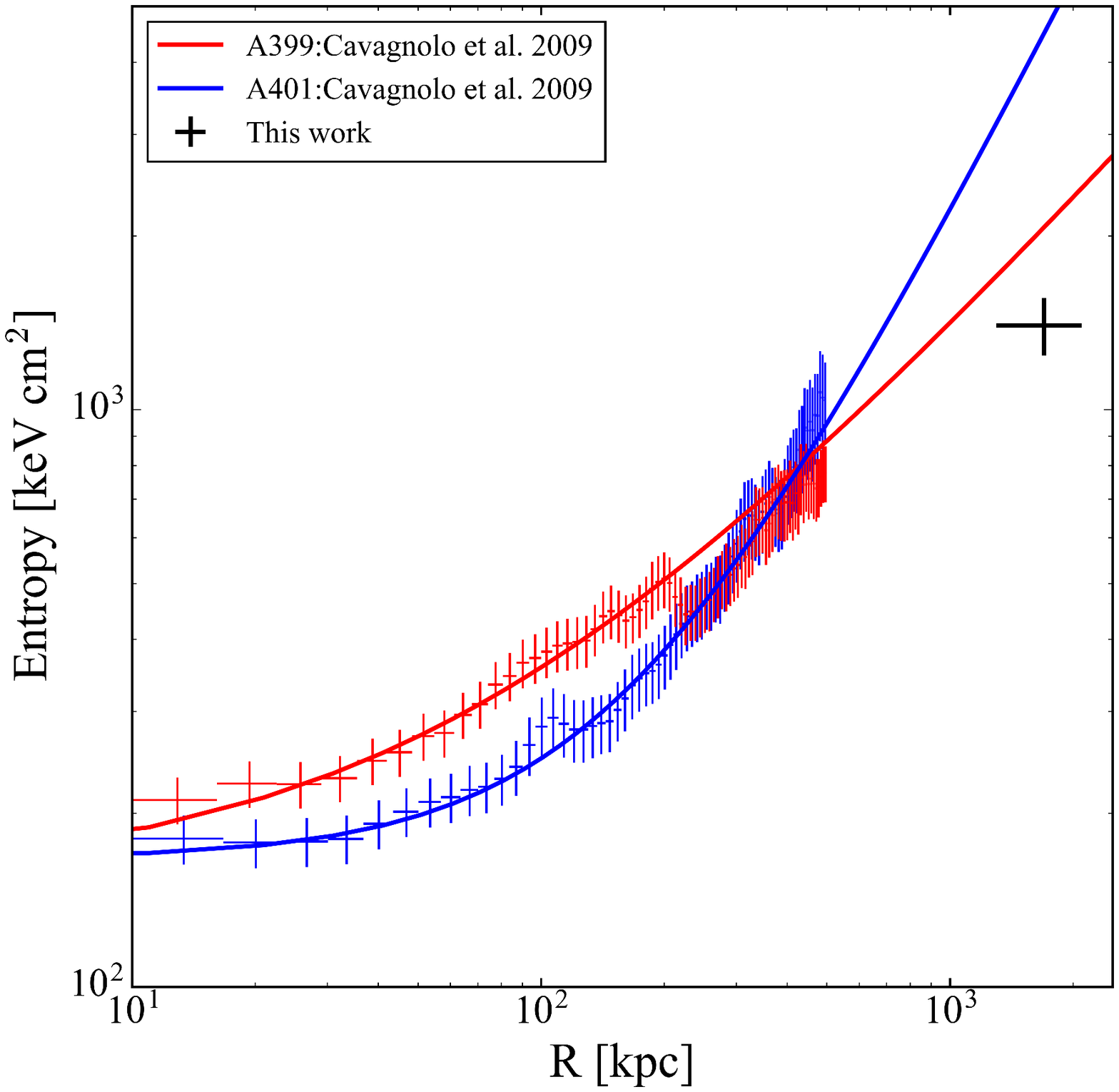}
\end{center}
\end{minipage}
\end{tabular}
\caption{\label{fig:property}
Left: Radial profiles of the pseudo-electron density, thermal pressure,  and entropy.  In the top panel, the orange region indicates the electron density from the {\it Planck} result of \citet{planck13_filament}. 
Right: Radial entropy profiles for A399 (red) and A401, respectively. Crosses show the results from {\it Chandra} observations~\citep{cavagnolo09}. Solid line represent the empirical model: $K(r)=K_{0}+K_{100}(r/100~\rm kpc)^{\alpha}$. We adopt the best-fit parameters from ~\citep{cavagnolo09}:$K_{0:A399}=153.20,
 K_{100:A399}=204.34, 
 \alpha_{A399}=0.79$, 
 $K_{0:A401}=166.80, 
 K_{100:A401}=81.79, 
 \alpha_{A401}=1.40$, respectively.
}
\end{figure*}

\section{Discussion}
{\it Suzaku} performed a deep observation of the filament plasma between A399 and A401. 
Thanks to low and stable instrumental background and the high sensitivity to diffuse sources of the {\it Suzaku} XIS, we were able to detect X-ray emission up to 1.3 Mpc from the collision axis.  In the next subsections, we briefly discuss the implication of these results.

\subsection{The properties and geometry of the filamentary plasma}\label{sec:property}
First let us estimate the properties of the filament plasma. We deduce an electron density in the filament from the {\it CIE} normalization. Assuming a volume $V$ of a long-box (1 arcmin$^2\times l$, $l$ is a line-of-sight depth of the volume) and a constant density within the assuming volume, the normalization of the {\textit cie} model is 
{\it Y}=$\int n_{e\rm} n_{\rm H} V\times 10 ^{64} \rm ~m^{-3}$, where, $n_{e}\sim 1.2 n_{\rm H}$ and {\it V} is the volume. 
Here $n_{\rm H}$ represents the Hydrogen density. 
From our analysis (Tab.~\ref{tab:results}), the electron density in isothermal regions  can be estimated as
\begin{equation}
n_{e}=n_{e0}~(\frac{l}{\rm Mpc})^{-0.5}~\rm cm^{-3},
\label{eq:ne}
\end{equation}
with $n_{e0}=2.5-3.3\times10^{-4}$. 
This value is broadly consistent with the previous estimation
\citep[$n_{e}\sim$ 3.4$\times$10$^{-4}~\rm cm^{-3}$, ][]{fujita08}.  Here, cosmological simulations suggest an exponential density profile perpendicular to the collision axis (see Introduction). 

The entropy ($\displaystyle K=kT/n_e^{2/3}$) of the ICM preserves the heating history of clusters. Therefore the entropy is a nice tracer of cluster growth~\citep{voit05}.  Combined with the measured temperature and electron density, we estimate the entropy of the filament. The estimated entropy is {\it K}$\sim$1300 keV~cm$^2$. Because of heating by accretion shocks, the entropy of the ICM increases towards the cluster outskirts. Previous {\it XMM-Newton} and {\it Chandra} observations revealed that the entropy of the ICM well exceeds $K$ $>$1000 keV~cm$^2$  above 1 Mpc~\citep{pratt10, cavagnolo09}. Our measurement covers up to the virial radius for each cluster.  At the cluster outskirts, the entropy is expected to show a much higher value. The right panel of Fig.~\ref{fig:property} shows the radial entropy profile of A399 (red) and A401 (blue), respectively~\citep{cavagnolo09}. Our measured entropy range is shown by the black cross. The measured low entropy indicates that the filament has not reached hydrodynamic equilibrium. In other words, the filamentary plasma is still not part of the clusters.
However, {\textit Suzaku} observations of clusters revealed that the entropy profiles tend to show lowers value than observed from the simulations~\citep[e.g.,][]{walker12_entropy}.
To make a firm conclusion, mapping observations are strongly desired.

Another important feature of the filamentary plasma is  the report of a tSZ effect signal by {\it Planck}~\citep{planck13_filament}. 
Because of the different dependencies on the temperature and electron density, the combination of the Compton {\textit y}-parameter and X-ray surface brightness allow us to evaluate the electron density and the line of sight length.
The {\it y}-parameter can be described as follows~\citep{sz72}:
\begin{equation}
y\equiv\int\frac{k_{\rm B}T_{\rm e}}{m_ec^2}n_e\sigma_T dl,
\end{equation}
where {\it k$_{\rm B}$} is the Boltzmann constant, $T_{\rm e}$ is the electron temperature, $\sigma_{\rm T}$ is the Thomson cross-section, $n_{e}$ is the electron number density,  $m_{\rm e}c^2$ is the electron rest mass energy, and the integration is along the line of sight. 
Here, let us estimate  the Compton {\it y}-parameter from the {\it Suzaku} results. 
Using {\it kT}=6.5$\pm$0.5 keV and $\displaystyle{n_{\rm e}=3.0\pm0.3\times10^{-4}(\frac{l}{\rm Mpc})^{-0.5}~\rm cm^{-3}}$, the estimated {\it y}-parameter is 
$\displaystyle{14.5\pm1.8 (\frac{{\textit l}}{\rm Mpc})^{0.5}\times10^{-6}}$. 
Compared with the result of {\it Planck}~\citep[14-17$\times10^{-6}$: see Fig. 2 in][]{planck13_filament}, 
the line of sight length and the electron density were estimated to be $l=1.1$ Mpc and $n_{\rm e}=3.1\times10^{-4}~\rm cm^{-3}$, respectively.
Thanks to its stable and low background level, {\it Suzaku} successfully detected X-ray emission from the filamentary structure up to {\it r}$\sim$15\arcmin~(1.3 Mpc) from the connecting axis. 
Assuming that the other side of the filament also has a similar scale,  the filamentary structure may span $\sim$2.6~Mpc with a depth of 1.1 Mpc.  This indicates that the morphology of the filamentary structure is likely  a pan-cake shape rather than a cylindrical shape. The combination of future high-sensitivity SZ and X-ray observations allows us to perform a better 3D tomography of cosmic filaments.
Assuming a volume of 2.6$\times$1.1$\times$1.0 Mpc$^{3}$ and a constant density of $n_{\rm e}=3.1\times10^{-4}~\rm cm^{-3}$, we obtain the gas mass of {\it M}$_{\rm gas}\sim$1.4$\times$10$^{13}$~{M}$_{\odot}$. 
We stress, however, that the estimated gas mass includes some contributions from both clusters. Furthermore, the assumption of a uniform density is a strong simplification.  Therefore, the quoted gas mass has to be considered as an upper limit.
Assuming the cosmic baryon fraction of 18\%~\citep{komatsu11}, the expected total mass of the filament is 7.7$\times10^{13}~\rm M_{\odot}$, which corresponds to about 10\% of the total mass of A399 and A401, respectively. We refer to the total mass of A399 and A401 from~\citet{sakelliou04} as 5$\times10^{14}~\rm M_{\odot}$ and 6.1$\times10^{14}~\rm M_{\odot}$, respectively.

\begin{figure}
\begin{tabular}{c}
\begin{minipage}{1\hsize}
\begin{center}
\includegraphics[width=1.\hsize]{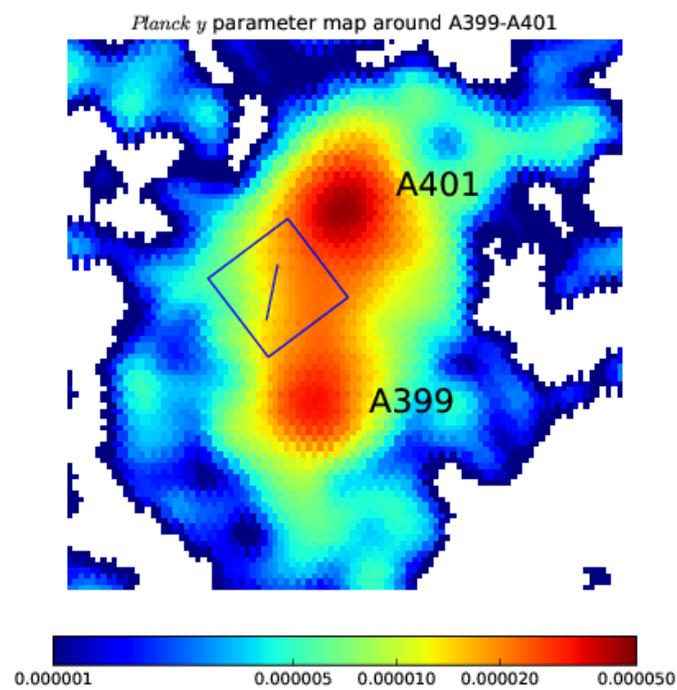}
\end{center}
\end{minipage}
\end{tabular}
\caption{\label{fig:y-parameter}
The Compton {\it y}-parameter around A399--A401 from \citet{planck16}. Blue box and line represent the {\it Suzaku} FOV and the location of the temperature break from the isothermal region.
}
\end{figure}

\subsection{The abundances at the cluster outskirts}
Using the same observation (and, therefore, within the same field of view as shown in Fig. 1), \citet{fujita08} measured a significant metallicity value of $\sim$0.2 $Z_\sun$ in two large bins ($\sim$8 arcmin broad each) out to the virial radii of 
both clusters. They concluded that the presence of these metals at large radii bear witness to powerful galactic superwinds having contributed to enriching the forming ICM at high redshift, i.e. before the clusters started to assemble. Since Fe is predominantly produced by Type Ia supernovae \citep[SNIa; e.g. ][]{deplaa07,mernier16b}, it is clear that a significant fraction of SNIa having contributed to the ICM enrichment must have exploded early on, i.e. before or when those galactic superwinds occurred. A similar result, also interpreted as the signature of an early enrichment by SNIa, has been found in the outskirts of the Perseus cluster~\citep{werner13}. This also suggests that the positive Fe abundance previously found in the outskirts of A\,399-A\,401 is likely not related to the particular configuration of these two systems (pre-merging phase).
\\
However, the two measurements alone in the analysis of \citet{fujita08} left some remaining doubt regarding the uniformity of the Fe abundance. From our reanalysis, and as shown by Fig. 4 (middle panel), it clearly appears that the Fe distribution remains remarkably uniform at smaller scales ($\sim$200 kpc) and within error bars.
\\
That said, care must be taken when interpreting metallicities in cluster outskirts. Even assuming (very optimistically) that the NXB and CXB components affecting spectra of low signal to noise regions are perfectly understood and under control, the moderate PSF of the XIS instrument prevents us from checking the spatial uniformity of metals below 200 kpc. In particular, \citet{molendi16} showed that, using current CCD spectrometers, the presence of unresolved cooler, denser, and more metal-rich substructures in the outskirts would significantly bias high the actual mass-weighted metallicity (typically by a factor of $\sim$3). Similarly, the limited quality of our spectra does not allow us to probe the thermal structure of the ICM at such a large distance from the core. Since the equivalent widths of the metal lines strongly depend on the plasma temperature, an inappropriate single-temperature modelling would affect the abundance measurements in case of a non-isothermal ICM~\citep[see e.g.][]{fujita96_abundance, fujita97, buote00}
\\
Keeping these uncertainties in mind, it must be noted (Fig. 3, bottom) that our average Fe abundance measurement of $\sim$0.2 $Z_\sun$~\citep[thus consistent with the value of][]{fujita08} agrees very well with the extrapolation of the radial Fe abundance profiles in clusters derived by \citet{matsushita11} and \citet{mernier17}. Assuming, as suggested above, that the Fe content of our investigated region is independent of the actual merging state of A\,399-A\,401, our results bring one more hint toward (at least) two distinct steps in the ICM enrichment history: (i) an early enrichment, presumably via galactic superwinds before the cluster assembly, responsible for a uniform Fe floor (see above); and (ii) in case of cool-core clusters, a later enrichment, likely due to the central cD galaxy, responsible for the central Fe peak\footnote{In principle, ram-pressure stripping of infalling cluster galaxies could also produce a centrally peaked Fe distribution \citep{domainko06}; however its predicted extent ($\sim$1 Mpc) is considerably broader than what is actually observed~\citep[e.g.][]{simionescu09}}. 
\\
Besides Fe, the  O, Mg, Si, and S abundances can in principle be measured with {\textit Suzaku} \citep[e.g.][]{sato07}. Since these four elements are synthesized (at least partly) by core-collapse supernovae (SNcc), measuring their abundances at large radii can provide us valuable information on the role of SNcc in the ICM enrichment history~\citep[e.g.][]{simionescu15, ezer17}. Unfortunately, the limited exposure of our data, together with the moderate spectral resolution of XIS and the high temperature plasma (giving rise to weaker metal line emissivities), prevent us from constraining the O, Mg, Si and S abundances in the region studied here. Future missions will be particularly useful to tackle this issue.

\subsection{The edge and origin of the filamentary hot plasma}~\label{sec:origin}
{\it Suzaku}'s high sensitivity and low background enable us to estimate the temperature structure of the filament with a high accuracy.   
In the region between  both  clusters, we find a clear  enhancement in the temperature profile of ICM from {\it kT}$\sim$4 keV (expected from the universal profile) to $\sim$6.5 keV (Fig.~\ref{fig:T_comp}).   This indicates that the
   clusters have been interacting even though the accrual distance between the
    clusters ($\sim$7 Mpc assuming  the redshift difference between the clusters is only due to the Hubble flow) is
 larger than their virial radii ($\sim$2 Mpc). 
Similar temperature enhancements were also reported in the A1750, Cygnus A, CIZA J1358.9-4750 and 1E2216.0-0401 system~\citep{belsole04, molnar13, sarazin13, kato15, akamatsu16}. 
For all those cases each of the two clusters are is partially located within the virial radius of its respective companion.
That is, it is natural to think that the cluster pair already starts interacting and most probably the temperature enhancement is caused by a compression induced by the interaction. On the other hand, in the A3528 case, there is no clear evidence of the enhancement in both the temperature and surface brightness~\citep{gastaldello03}. A similar result has  also been reported in the A3556~\&~A3558 system~\citep{mitsuishi11}. 
That is, in the case where the separation of  the pair is much larger than their virial radius, the temperature enhancement  is a not so common phenomena.

Our scenario is that  A399\&A401 originally had a filament connecting the clusters, which caused an accretion flow onto the filament and formed a shock front parallel to the connecting axis.
During the evolution of the clusters, the filament is (adiabatically) compressed by the merging activity. This increases the gravitational force of the filament. Consequently, the filament should have a strong accretion flow onto it. In this case, the observed filament plasma could not be the ICM of the clusters only, most likely  it is a mixture of the ICM of the clusters and intergalactic medium (IGM), which was
 not captured by the clusters yet. 
As we discuss in the next sections (Sect.\ref{sec:comp_sim}),  predictions of cosmological hydrodynamical simulations~\citep{dolag06} of a filament located between two clusters match these measured properties of our studied filament candidate, such as uniform temperature, {\it y} parameter within the filament and 
a possible shock front parallel to the connecting axis.
Furthermore, we investigated a relationship between the {\it y} parameter measured by {\it Planck} and the observed temperature break. The bottom panel of fig.~\ref{fig:y-parameter} shows the {\it y} parameter map reported by the~\citet{planck16}.

Based on the assumption that the temperature break is a shock front, we will briefly discuss the implication of these results. To evaluate the intensity of the shock, the Mach number,  we applied the Rankine-Hugoniot jump condition,
\begin{equation*}
{\frac{T_2}{T_1}=\frac{5{\cal M}^4+14{\cal M}^2-3}{16{\cal M}^2}},
\end{equation*}
where subscripts 1 and 2 denote pre-shock and post-shock values, respectively.
Here, we assume the post and pre shock temperature as $kT_{\rm post}\sim6.3~ {\rm keV}, kT_{\rm pre}\sim5.1~\rm keV$ (Table~\ref{tab:shock}). The resultant Mach number is about ${\cal M}\sim1.3$. 
These results support the interpretation that the shock that we are witnessing  is associated with the rim of the filament, which was formed by the accretion flow due to the gravitational force by the filament. 

Comparing with recent findings of a shock front in an early phase merging cluster~\citep{kato15, akamatsu16}, the property (the Mach number) of the shocks is almost consistent with each other. The location of these shocks are well matched with the prediction of numerical simulations of major merger events~\citep{takizawa99,ricker01}.
Contrary to the other samples, we observed the shock front at a different geometry.  
Previous hydrodynamical simulations of A399 and A401 without the contribution of the filament~\citep{akahori08} also failed to predict such a shock front with a pure binary cluster merger. XMM-Newton observations revealed that the surface brightness profile between A399 and A401 can not be explained by just simple superposition of the two clusters \citep[see the right panel of Fig.10 in ][]{sakelliou04}. Furthermore, our work suggests there is more mass than the simple superposition. These observational facts support the presence of a considerable amount of WHIM filament between the clusters at initial phase.

\subsection{Comparison with the hydrodynamical simulations}\label{sec:comp_sim}
The time scale of the evolution of such large-scale filaments and clusters of galaxies is compatible with the Hubble time. In this case, there are two approaches to understanding their  full  evolution history. The first approach is making snapshots of the history by collecting several samples which are at a different evolution phase. Another approach is a comparison with numerical simulations, which is able to track the full evolution of such large scale structures with already known physics.  

In practice, the number of observed cases is not very large.
Thus, we compare our results with previously performed numerical simulations. We summarised our results as follows; 
\begin{enumerate}
\item A nearly isothermal ({\it kT}$\sim$ 6--7 keV) filament between two clusters of galaxies: Sect. \ref{sec:property} and Fig:~\ref{fig:T_pro}
\item The Compton {\it y} parameter of the tSZ effect is
 $\sim 1.5\times10^{-5}$: Sect.\ref{sec:origin} and Fig.~\ref{fig:y-parameter}
\item The possible presence of a (weak) shock at the outer part of the filament: Sect.\ref{sec:origin} and Fig~\ref{fig:T_pro}
\end{enumerate}

We compare these results with the simulations presented in \citet{dolag06}. 
The simulations concentrate on describing the physical properties of a supercluster-like region including the treatment of radiative cooling, heating by a uniform UV background and star formation feedback processes. For the comparison, we selected the cluster pair named as halo-b and halo-d (See Table 1 for the details). 
This cluster pair reasonably mimics the situation of A399--A401. At {\it z}=0.07, the physical separation of the pair is about 4 Mpc/{\it h} (3.0 Mpc/{\it h} in the projection on the sky plane).
The main features of halo b-d can be summarised as follows:
\begin{enumerate}
\item Accretion flows onto the filament, which generates shock fronts parallel to the axis joining the two haloes: Sect.3 in their paper
\item Due to the outward movement of the accretion shocks are created
 extended, almost isothermal regions ({\it kT}$>$ 4.3 keV): Sect.4 and the middle of the left panel of Fig.6
\item The tSZ Compton-y parameter can be the order of $\sim10^{-5}$: Fig.10.
\end{enumerate}

These physical properties of cosmic filaments predicted by the simulation~\citep{dolag06} are, qualitative speaking, in an excellent agreement with our measurements of the filamentary plasma between A399--A401. Therefore, the observed temperature break is likely the shock front related to the structure formation, which is still propagating  outwards together with ongoing evolution of the filament. 

\section*{Acknowledgments}
The authors thank the referee for the constructive comments which have helped to improved the paper.
The authors would like to thank all the members of the {\it Suzaku} team for their continuous contributions in the maintenance of onboard instruments, calibrations, software development.
H.A acknowledges the support of NWO via a Veni grant. YF acknowledges financial support from MEXT KAKENHI No.15K05080.
SRON is supported financially by NWO, the Netherlands Organization for Scientific Research. 

\bibliographystyle{aa}
\bibliography{references}
\appendix
\label{lastpage}

\end{document}